\documentclass[useAMS, usenatbib]{mn2e}

\usepackage{graphicx}
\usepackage{aas_macros}
\usepackage{times}
\title[Dust Scattering in Miras R Car and RR Sco]
{Dust Scattering in Miras R Car and RR Sco resolved by 
optical interferometric polarimetry}

\author[M.J. Ireland et al.]{M.J. Ireland, P.G. Tuthill, J. Davis, W. Tango\\ 
School of Physics, University of Sydney NSW 2006, Australia}

\begin{document}

\pagerange{\pageref{firstpage}--\pageref{lastpage}} \pubyear{2004}
 
\maketitle

\label{firstpage}

\begin{abstract}
We present optical interferometric polarimetry measurements of the Mira-like 
variables R~Car and RR~Sco, using the Sydney University Stellar Interferometer. 
By making visibility measurements in two perpendicular polarisations, the 
relatively low-surface brightness light scattered by atmospheric dust
could be spatially separated from the bright Mira photospheric flux.
This is the first reported successful use of long-baseline optical 
interferometric polarimetry.
Observations were able to place constraints on the distribution of 
circumstellar material in R Car and RR Sco.
The inner radius of dust formation for both stars was found to be less than 
3 stellar radii: much closer than the expected innermost stable 
location for commonly-assumed astrophysical ``dirty silicate'' dust in these 
systems (silicate dust with a significant iron content). 
A model with the dust distributed over a shell which is geometrically thin
compared to the stellar radius was preferred over an outflow. 
We propose dust components whose chemistry and opacity properties enable 
survival at these extreme inner radii.
\end{abstract}


\section{Introduction}

The processes of dust formation and mass loss in evolved giants and 
supergiants remains poorly understood despite its manifold and profound 
implications to the cycle of matter through stars and back into the ISM. 
An extensive observational and phenomenological literature (see e.g.
\citet{Habing96} for a review) bears witness to the barrage of astrophysical
techniques applied to this problem, and progress has been considerable in many 
important areas.
However, fundamental uncertainties remain over the mechanisms of elevation 
of material, chemical nucleation of the dust, and subsequent outflow and
global properties of the wind. 
As an example of the gap in our understanding, there has been to date
no self-consistent physical model of the atmosphere of an evolved
O-rich giant which encompasses the upper regions including even
the most important basic phenomena:  non-grey radiative transfer,
pulsation, shocks, convection, non-equilibrium dust formation
and the molecular atmosphere.

From an observational perspective, progress has been hampered until now by 
the difficulties inherent in imaging material on size scales of a few AU 
which encompass the photospheres themselves and the dust formation and 
acceleration regions.
Although modern long-baseline optical/IR interferometry is capable of
accessing these scales, imaging is not straightforward as targets can be 
complex, deeply embedded and often asymmetric.
In particular for many Mira-type variables, the relatively low 
surface-brightness material in the outflow can be very difficult to 
detect against the high luminosity stellar flux.

The stellar photosphere and circumstellar dust shell are usually 
treated as completely separate regimes, as justified by the following
simple argument.
As a first
approximation, one might consider that dust grains in an
  optically thin envelope are grey or nearly
so, and exist in thermal equilibrium with the radiation field. 
Furthermore, we can approximate the star as having a small
  angular extent, so that the mean intensity, $J(\lambda)$ at the dust
  condensation radius $R_c$ is
  equal to the luminosity of the central star $L(\lambda)$
  divided by $16\pi^2R_c^2$. In this case
Equation~\ref{eqnDustGrey} gives an estimate of the dust formation
radius (here $R_s$ the radius
of the central star, $T_s$ the effective temperature of the star and
$T_c$ the dust condensation temperature). Given an effective temperature 
of 3000\,K and a dust condensation temperature of 1100\,K, the
dust formation radius is 3.7 stellar
radii, which means that
the physics of the photosphere and the circumstellar environment could
be considered well-separated. Where dust is considered to be non-grey,
but the dust formation radius still lies well-outside the photosphere, 
conservation of energy and radiative equilibrium gives
Equation~\ref{eqnDustAway}, where $Q$ is the
absorption coefficient of the dust and $B$ is the Planck
function in the form which gives power per unit area per steradian 
per unit wavelength. Further discussion of these kinds of
  approximations can be found in \citet{Schutte89}.

\begin{equation}
 R_c = \frac{R_s}{2}(\frac{T_s}{T_c})^2
\label{eqnDustGrey}
\end{equation}

\begin{equation}
 R_c = \sqrt{\frac{\int Q_{\rm abs}(\lambda) L(\lambda) d\lambda} 
                  {16\pi^2\int Q_{\rm abs}(\lambda) B(\lambda, T_c) d\lambda}}
\label{eqnDustAway}
\end{equation}

The absorption coefficients for dust around O-rich Mira are
often assumed to be those of ``dirty'' silicates \citep{Jones76, Ossenkopf92}. 
These silicates have optical constants very similar to amorphous
olivine Mg$_x$Fe$_{1-x}$SiO$_4$ with $x=0.5$ \citep{Dorschner95}, with
the exception of the spectral range 2-8\,$\mu$m, where inclusions
such as solid Fe are required to match observationally derived optical
constants. Olivine with $x=0.5$ can form only slightly closer to the star than
grey dust. Using the emitted flux from a 3000\,K effective temperature
Mira model from Ireland et al. (2004a) (the M09n model), the optical constants from
\citet{Dorschner95} and an assumed condensation temperature of
1100\,K, this olivine can form at a radius of 6.4 stellar radii: 
still a comfortable distance from the upper atmosphere of the star.

Apart from the optical properties, the other factor governing grain 
survival at smaller radii is the sublimation temperature. The
dust species often considered to be the first to form in higher
effective temperature stars from semi-empirical considerations is
corundum \citep{Egan01}, Al$_2$O$_3$,
which is stable at considerably higher temperatures than silicates
\citep{Salpeter77}. Its
formation radius for this same 3000\,K model star is 2.1 stellar radii,
using the optical constants from \citet{Koike95} and an
assumed condensation temperature of 1400K.

The multi-wavelength diameter measurements of six Mira-like variables made
by Ireland et al. (2004b) show an increase in apparent diameter from
900\,nm to 700\,nm with a component which appears uncorrelated
with the strength of TiO bands. These observations are best explained
by scattering from dust closer than a few stellar radii
(although molecular line blanketing makes quantitative interpretation
difficult).
The presence of extensive dust this close is, as discussed above, difficult 
to reconcile with the conventional wisdom on dust formation around AGB stars.

Despite the fact that the net polarisation arising from a spherically
symmetric dust shell is zero, it is possible to separate it from the
unpolarised photosphere with a polarisation-sensitive high-resolution
measurement.
Scattering from the dust shell will have a polarisation at any 
point with the E field in an azimuthal direction with respect to the
stellar centre. This is due to the simple relationship that scattered light from
small particles in the Rayleigh limit is polarised orthogonal to
the plane containing the source, scatterer and observer.

If we now consider the response of a long-baseline interferometer
which measures the Fourier components (called visibilities when total 
flux is normalised to unity) of such an image, these will have different 
amplitudes and phases in different polarisations. 
For a partially-resolved circularly symmetric source, the linear 
polarisation with the E field parallel to the baseline of observation
will have a lower amplitude and the same phase as the visibility
measured using the polarisation perpendicular to the baseline.
An attempt was made using this technique
to detect scattering by free electrons around $\beta$ Orionis using
the long-baseline Narrabri Stellar Intensity Interferometer
\citep{Brown74}, but the instrument lacked the required sensitivity.
Here we report the first successful use of this technique on a 
long-baseline optical interferometer. 
The observational technique is described in the following section, 
while Section~\ref{sectModel} gives the results and Section~\ref{sectDiscuss} 
a discussion and astrophysical interpretation.

\section{Observations}
\label{sec:obs} 

The Sydney University Stellar Interferometer (SUSI) is a long baseline
optical interferometer with presently operational baselines from 5 to
160\,m (further details can be found in \citet{Davis99}). 
The observations presented here used 5 and 10\,m baselines
and a filter with central wavelength of 900\,nm and full-width
half-maximum 80\,nm. This filter was chosen so as to minimise
contamination from the strongest features caused by the TiO
molecule. It is in principle possible to measure visibilities at arbitrary
polarisations through the use of quarter-wave plates and polarisers at
arbitrary orientations. For a discussion of a general way to calibrate
a long baseline interferometer for imaging in all Stokes parameters,
see \citet{Elias04}. Due to polarisation-dependent phase-shifts and
mirror reflectivities over the many mirrors in the SUSI optical chain,
arbitrary initial polarisation states are not well preserved.
The measurements therefore are restricted to states with pure vertical 
and horizontal electric field orientation with respect to the optical 
path.
While they will be attenuated, these states should not suffer mixing
from any reflection with the exception of the initial reflection off
the siderostat when observing away from the meridian. Care was taken
to only observe stellar targets when near the meridian. 
Given SUSI's North-South baseline orientation, this observing strategy 
resulted in visibility measurements with the E field at all times being
within 19 degrees of perpendicular ($V_{\perp}$) or parallel ($V_{\|}$) 
to the baseline.
For crucial observations where a direct comparison was made between
the visibilities in these two polarisations, data were restricted further
to lie within 8 degrees of these orientations.

\begin{figure}
 \includegraphics{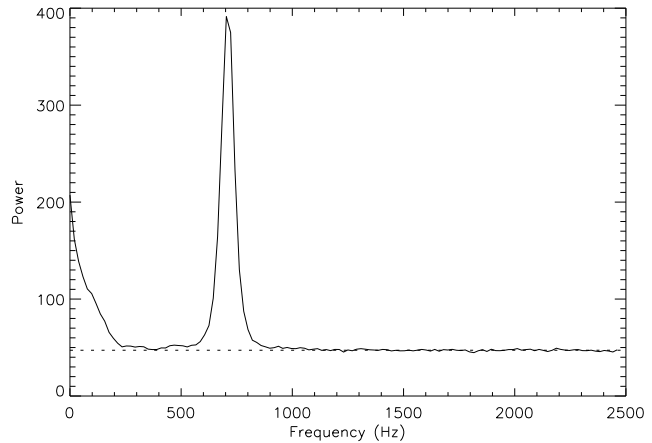}
 \caption{An example power spectrum (for the calibrator star HR~6241
 in the 900\,nm filter)
 obtained from averaging the power spectra from 3000 fringe scans
 together. Power is in arbitrary units. Note that although the
 scintillation power is clear at low frequencies, bias subtraction is
 relatively simple, and in this case a constant bias (dashed line)
 would be a good approximation over the range of frequencies covered by the 
 fringe peak.}
 \label{figPowerSpectra}
\end{figure}

The beam combining system used for these observations consisted of a
pupil-plane beam combiner feeding two Avalanche Photo-Diode (APD) detectors. 
By repeatedly scanning the optical delay linearly through the 
white-light fringe position, the light detected from each output of
the combining beamsplitter is modulated at a frequency that depends on the 
scanning rate and the observing wavelength, with the fringe signals
180 degrees out of phase at the two outputs. A description of a
similar system can be found in \citet{Baldwin94}, and further
details of the beam combining system used for these observations
will be found in \citet{Davis05} (in preparation). 
Although the two outputs are usually used differentially to reduce the 
effects of scintillation noise, 
it is possible to obtain an estimate for the fringe visibility 
from only one output, as in Figure~\ref{figPowerSpectra}. 
The use of only one detector means that the bias in the power spectrum is not 
flat due to scintillation, and incorrect subtraction of this bias becomes an
additional error source for low $V^2$.
The observational strategy employed consisted of two separate parts: 
the first aimed at measuring $V_{\perp}$ and $V_{\|}$ separately, while
the second measured the ratio $V_{\perp}/V_{\|}$. 
These are described in turn below.

Firstly, polarisers were co-aligned and placed in front of the two outputs 
of the beam combiner, enabling a high signal-to-noise measurement of the 
correlation in linear polarised light.
As is usual with optical interferometry measurements, observations of 
targets were interleaved with un- or partially-resolved reference stars
to calibrate the system correlation (the $V^2$ response to a point source)
as a function of time.
Before use in adjusting the science measurements, this system response was 
also corrected for the known nonzero diameters of the reference stars
given in Table~\ref{tblCalStars}. 
In this phase of the observational program, measurements were predominantly
made of $V_{\perp}$ which had a much higher signal-to-noise than $V_{\|}$.
This is mainly due to partially polarising beamsplitters used at SUSI to 
split off light to the tip/tilt adaptive optics camera, which had the effect
of significantly lowering the system throughput for $V_{\|}$.

\begin{table}
 \caption{Assumed diameters for the calibrator stars used for these observations}
 \begin{tabular}{@{}lll@{}} 
 \hline
  Star & Uniform Disk (UD)   & Method $^2$\\
       & Diameter (mas)      &            \\         
 \hline
  HR 6241 & 5.9     & VB99 \\
  HR 6630 & 4.2     & VB99 \\
  HR 4050 & 4.9     & C99  \\
  HR 6553 & 2.2$^1$ & O82  \\
  HR 3685 & 1.54    & O82  \\
  HR 3884 & 2.5,2.8 & S05  \\
 \hline
 \end{tabular}
\newline
 $^1$~This star also has a companion 3.39 magnitudes fainter and at
distance of 6.47$\arcsec$ (figures taken from the {\em HIPPARCOS}
catalogue). This was over-resolved by SUSI and taken into
account in the calibrated values for $V^2$.\newline
 $^2$~VB99: Diameters estimated from the B-K relationship from
 \citet{vanBelle99a}, using the photometry from the catalogue
 of \cite{Ducati02}. \newline
 C99: Diameter taken from \citet{Cohen99}, with limb-darkening
 correction from \citet{Claret00}. \newline
 O82: Diameters taken from \citet{Ochsenbein82}. \newline
 S05: Diameters of this Cepheid taken from near-simultaneous
 measurements at SUSI, at early April and early May epochs, consistent
 with the published diameters of \citet{Kervella04}.
 \label{tblCalStars}
\end{table}
%
%

The second part of the observing strategy consisted of simultaneous 
observations of $V_{\perp}$ and $V_{\|}$, made possible
by placing polarisers with orthogonal orientations in front of the two
outputs of the beam combiner. 
These observations gave a measurement of the ratio $V_{\perp}/V_{\|}$,
and due to the simultaneous data, the ratio could be obtained independent 
of seeing effects so that rapidly interspersing calibrator star
observations was no longer necessary, and longer total integration
times were possible. 
Calibrator stars, observed at a different time of night, were still used 
for these observations to correct for instrumental effects that might affect
this visibility ratio. 
Some calibrators were moderately resolved, and calibrators of a range of
spectral types were used, however there was no evidence for systematic
variation of the visibility ratio $V_{\perp}/V_{\|}$ between calibrator stars. 

A typical observation in a single polarisation consisted of 4000 50\,ms
sweeps through the fringe envelope, while the simultaneous
observations of $V_{\perp}$ and $V_{\|}$ contained up to 20000 50\,ms
sweeps through the fringe envelope. The slew and acquisition time
between source and calibrator generally averaged 3-4 minutes.
Details of all observations can be found in Table~\ref{tblObservations}.
Note that squared visibility (or correlation) quantities are listed, 
because in the photon-noise limited regime, squared visibility must be
averaged over many scans and error distributions are closest to
Gaussian in these quantities.


There are several sources of systematic error in the measurement of
$V^2_{\|}/V^2_{\perp}$. One error source which was corrected for in the
values in Table~\ref{tblObservations} was a
5\,nm shift in effective centre wavelength between the filter 
in each output channel. The effect of this was to make the souce
  appear slightly more resolved in one channel then the other.  This
  effect was well characterised by observing in a mode where both
  output channels observed the same polarisation state, and resulted
in a maximum correction of 3\% 
in $V^2_{\|}/V^2_{\perp}$ for the most well-resolved observation.

Other errors, not included as their effects were computed to be
small, consisted of imperfect polariser alignment, offsets 
between the actual polarisation passed and the ideal desired sky 
orientation, and depolarisation and mixing due to alignment errors 
in the beam train. With the assumptions of spherical symmetry and
  rayleigh scattering, each of these effects multiplies $V_{\|} -
  V_{\perp}$ by a factor $\gamma$ which is has a modulus smaller than
  1. We will
  define $\phi$ to be the azimuthal coordinate describing the
  rotation of the siderostat normal about a horizontal north-south
  axis, with $\phi = 0$ describing the meridian. The effect of
  vertical/horizontal linear polarisation
  rotation with respect to the baseline orientation then gives a
  factor $\gamma_1 = \cos(2\phi)$. The mixing of linear with circular
  polarisation states due to phase changes on reflection gives a
  second factor $\gamma_2$. If we define $\delta$ to be
  the phase shift between $s$ and $p$ polarisations at the siderostat
  mirror (itself a function of the angle of incidence of the starlight), 
  then $\gamma_2$ is given by:

\begin{equation}
 \gamma_2 = \cos^2(\delta/2) + \sin^2(\delta/2)\cos(4\phi)
\end{equation}

For example, $\delta$ is 37 degrees for SUSI's 
siderostats at 900\,nm and an angle of incidence of 45 degrees,
which gives $\gamma=0.991$ for the maximum $\phi$ of 6~degrees for 
dual-polarisation measurements RR ~Sco. 
Taking all of these effects into account, we estimate that for no
 observation will the true
values for $V^2_{\|}/V^2_{\perp}$ be more than 0.6\% 
higher than the measured values given in Table~\ref{tblObservations}. This
is smaller than statistical measurement errors.

\begin{table*}
 \caption{Summary of SUSI observations of R Car and RR Sco. Visual phase (Col 4) were
  obtained from AAVSO data. The projected baseline $B_{p}$ and position angle range are
 given in Cols 5 \& 6, while the type of measurement and its recorded value are in 
 Cols 7 \& 8}
 \begin{tabular}{@{}lllllllll@{}}
 \hline
  Date & Star & JD & Phase & $B_{p}$ (m) & PA range ($^{\circ}$) & Measurement & Value \\
 \hline
   2004 Apr 8 & R Car & 2453104 & 0.08 & 4.2 & -15..-6 & $V^2_{\perp}$          & $0.582\pm0.04$ \\
   2004 Apr 8 & R Car & 2453104 & 0.08 & 8.4 &  1..18  & $V^2_{\perp}$          & $0.123\pm0.015$ \\
   2004 Apr 8 & R Car & 2453104 & 0.08 & 4.2 & -5..0   & $V^2_{\|}/V^2_{\perp}$ & $1.160\pm0.03$ \\
   2004 Apr 9 & R Car & 2453105 & 0.08 & 4.2 & -10..-2 & $V^2_{\perp}$          & $0.556\pm0.04$ \\
   2004 Apr 9 & R Car & 2453105 & 0.08 & 8.4 & 5..10   & $V^2_{\perp}$          & $0.125\pm0.015$ \\
   2004 Apr 9 & R Car & 2453105 & 0.08 & 8.4 & -1..4   & $V^2_{\|}/V^2_{\perp}$ & $1.220\pm0.06$ \\
   2004 May 1 & R Car & 2453127 & 0.15 & 4.2 & -12..0  & $V^2_{\perp}$          & $0.592\pm0.04$ \\
   2004 May 1 & R Car & 2453127 & 0.15 & 4.2 & 3..8    & $V^2_{\|}/V^2_{\perp}$ & $1.095\pm0.03$ \\
   2004 May 2 & R Car & 2453128 & 0.15 & 8.4 & -4..0   & $V^2_{\perp}$          & $0.129\pm0.01$ \\
   2004 May 2 & R Car & 2453128 & 0.15 & 8.4 &  -4..-1 & $V^2_{\|}/V^2_{\perp}$ & $1.150\pm0.11$ \\
   2004 May 6 & R Car & 2453132 & 0.16 & 8.4 & 0..5    & $V^2_{\|}/V^2_{\perp}$ & $1.220\pm0.10$ \\
   2004 Jul 28& RR Sco& 2453215 & 0.95 &10.0 & 4..7    & $V^2_{\perp}$          & $0.196\pm0.012$ \\
   2004 Jul 28& RR Sco& 2453215 & 0.95 &10.0 & -10..-7 & $V^2_{\|}$             & $0.259\pm0.015$ \\
   2004 Jul 28& RR Sco& 2453215 & 0.95 &10.0 & -5..2   & $V^2_{\|}/V^2_{\perp}$ & $1.195\pm0.05$ \\
   2004 Jul 29& RR Sco& 2453215 & 0.95 & 5.0 & -11..10 & $V^2_{\perp}$          & $0.646\pm0.01$ \\
   2004 Jul 29& RR Sco& 2453215 & 0.95 & 5.0 & -2..5   & $V^2_{\|}/V^2_{\perp}$ & $1.063\pm0.014$ \\
 \hline
 \end{tabular}
\label{tblObservations}
\end{table*}


\section{Results and Model fitting}
\label{sectModel}

\begin{figure*}
 \includegraphics{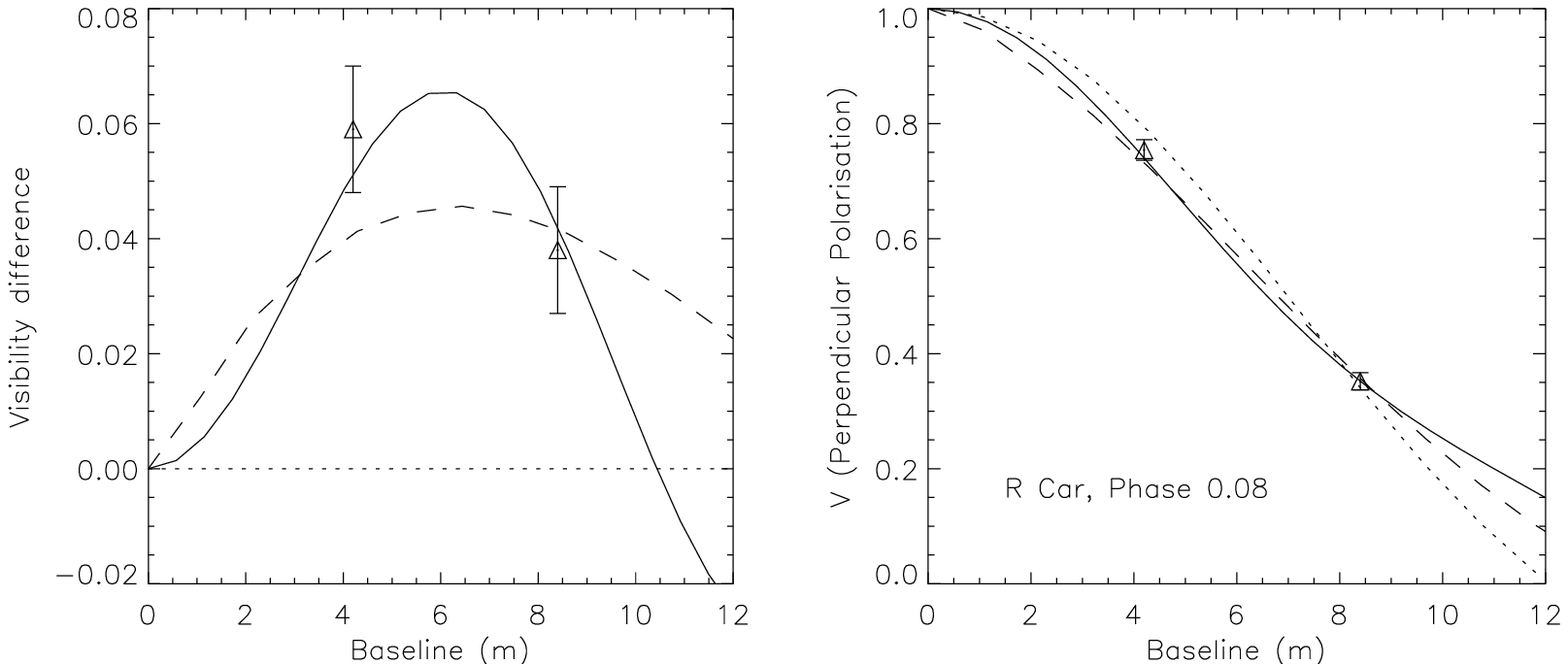}
 \includegraphics{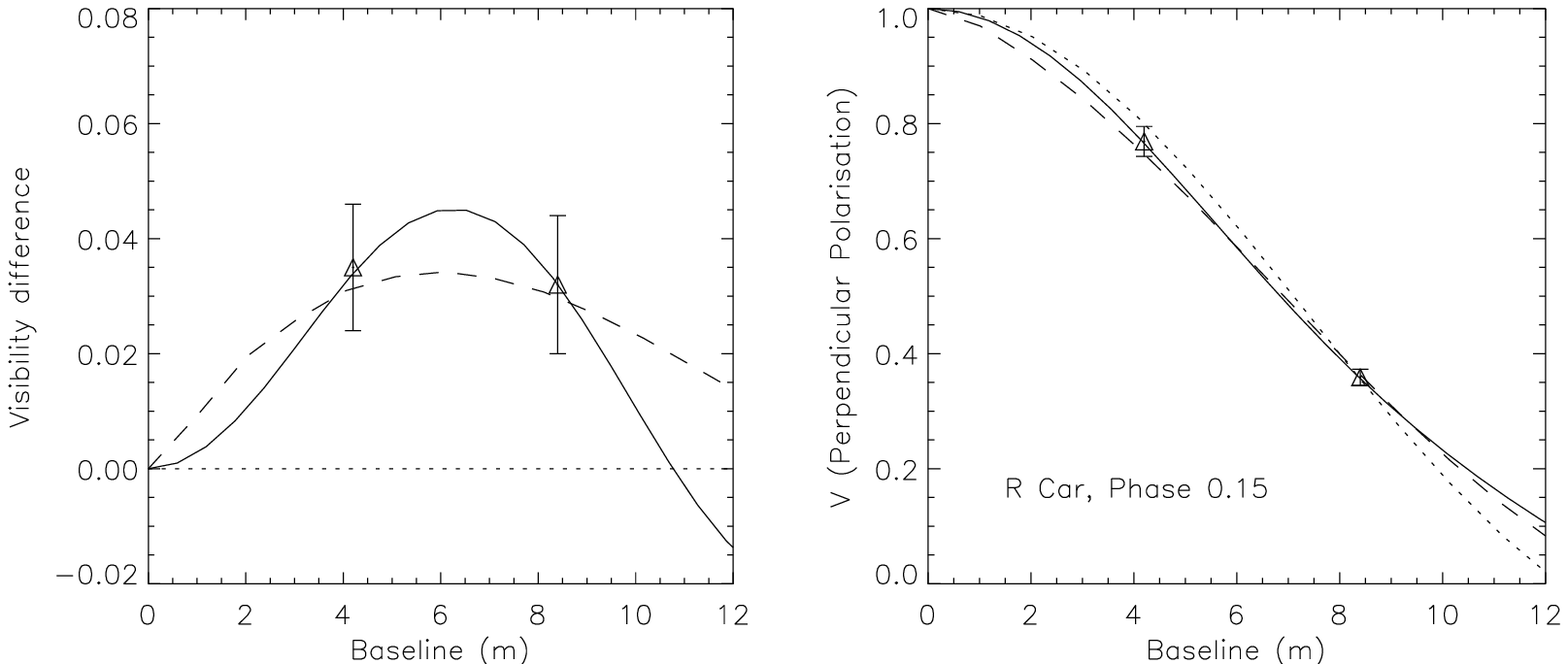}
 \includegraphics{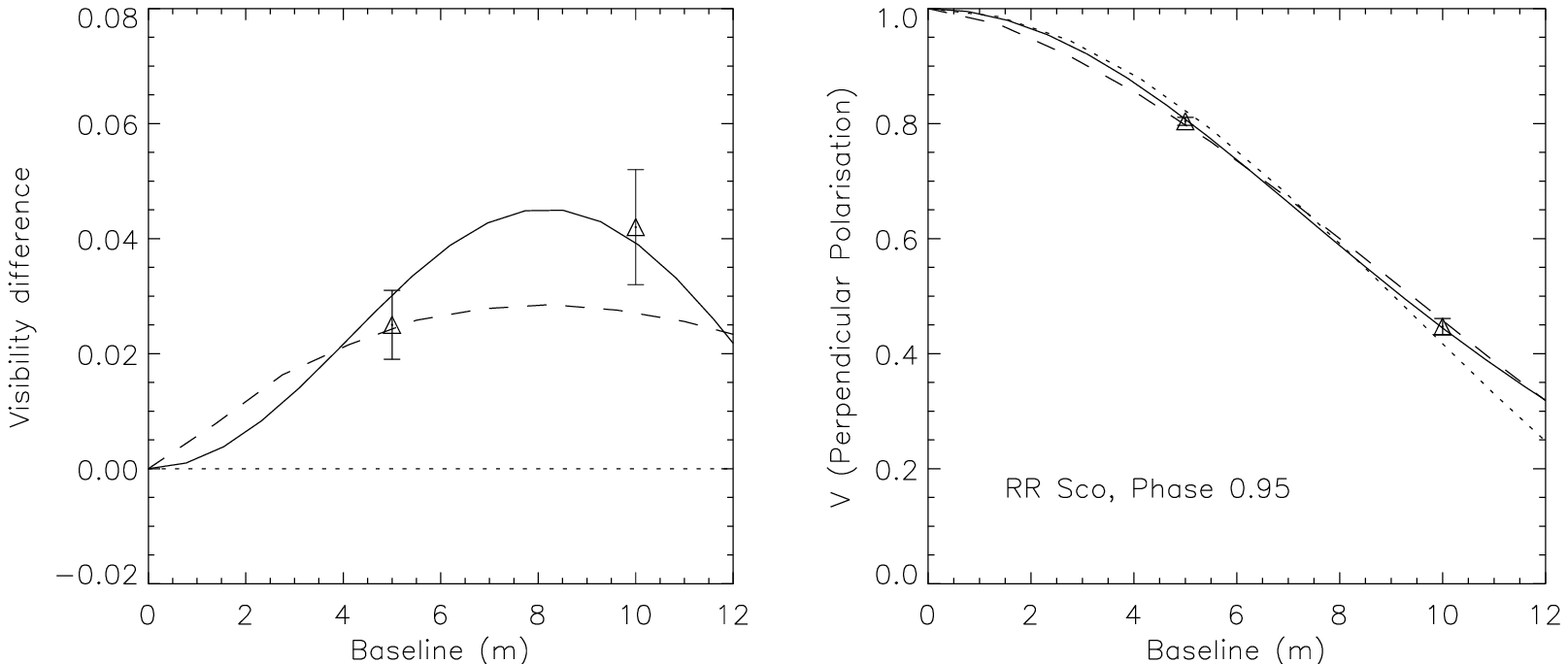}
 \caption{Fits to the first epoch (phase 0.08) of R~Car (upper plots),
 the second epoch (phase 0.15) of R~Car (middle plots) and the single
 epoch (phase 0.95) of RR~Sco
 (lower plots) with data reduced to $V_{\|}-V_{\perp}$ (left plots) and
 $V_{\perp}$ (right plots). The solid line is the thin shell model fit,
 the dashed line is the outflow model fit and the dotted line is the
 uniform disk only fit.}
 \label{figRCarFit}
\end{figure*}

As a first step in interpretation of the results, from
the observed quantities in Table~\ref{tblObservations} we
derive the quantities $V_{\|}-V_{\perp}$ and $V_{\perp}$.
For R~Car, we have averaged data into four points at projected baselines of 4.2
and 8.4\,m and covering pulsational phases of $\sim 0.08$ and $\sim 0.15$. 
Similarly, observations of RR~Sco have been distilled into measurements
at baselines of 5.0 and 10.0\,m with a phase of 0.95.
As the signal-to-noise of the initial observations is high, the
transformation from the observed quantities in $V^2$ to 
$V_{\|}-V_{\perp}$ and $V_{\perp}$ could be performed while preserving
the symmetric Gaussian error distributions reasonably well.

We assume that thermal photons from the star 
can be well represented by a Uniform Disk (UD)
visibility curve, and that the dust distribution is spherically
symmetric. The first assumption that the star is UD-like is reasonable
because at phases near maximum, the atmosphere of a Mira variable is
expected to be relatively compact (e.g. see Ireland et al. (2004a)), and
the filter was chosen so that there was minimal contamination by the 
TiO molecule. 
The second assumption that the dust is spherically symmetric is 
reasonable for R~Car because of its low total polarisation. Although
any measured polarisation may be interstellar, a low total
polarisation signal places a rough upper-limit on the amount of
light scattered by asymmetrical regions closest to the star.
Measured polarisation fractions are 0.62\% at 652\,nm
\citep{McLean77}, 0.52\% at 652.5\,nm \citep{CodinaLandaberry80} and
0.51\% in V band \citep{Serkowski01}, with all authors measuring a
significant downward trend in polarised fraction with increasing
wavelength. These numbers can be compared with the roughly 15\% total
scattered light as measured in this paper at 900\,nm. No such
comprehensive polarimetry measurements could be found in the
literature for RR~Sco.

Two dust models were fit to the data: a thin shell and an outflow.
For the thin shell model dust is assumed to exist in a
static infinitesimally thin shell, while for the outflow model dust 
forms at some condensation radius and is then radiatively accelerated
producing a stellar wind.
For the outflow model, the dust density as a function of 
radius can be derived as \citep{Schutte89}:

\begin{equation}
  \rho(r) \propto 1/[(r - (1-v_c^2))^{1/2}r^{3/2}]
\label{eqnOutflow}
\end{equation}

Here $r$ is radius in units of the condensation radius and $v_c$
is the outflow
velocity at $r_c$ as a fraction of the final
outflow velocity. We only consider the limiting case of $v_c=0$ in
this paper as it is provides a better fit to the data in all
cases. After fixing $v_c$ to zero, both models have only two free
parameters, the radius of dust formation 
and the optical depth of the dust shell. In both models, we assume that
the dust grain radii are much smaller than the observing wavelength,
so that the Rayleigh scattering approximation can be used, and the
light scattered by the dust is all assumed to originate from
the central star (i.e. only a single scattering is included). 

The polarisation signal from a dust shell illuminated by an extended
photosphere will be weaker than that illuminated by a compact object (or 
point source) of equivalent luminosity at the stellar centre.
We proceed here to derive the dilution of the signal due to the geometrical
extension of the underlying star.
Consider a dust grain at a distance $r_d$ from the centre of a star whose
radius is $r_s$.
The polarisation properties of light scattered from this grain will vary 
according to the angle of the incident photon, which in turn depends on 
the original location of the emission on the surface of the star.
The net polarisation is then found by integration over the disk of the
star visible to the dust grain.

\begin{figure}
\includegraphics[scale=0.7]{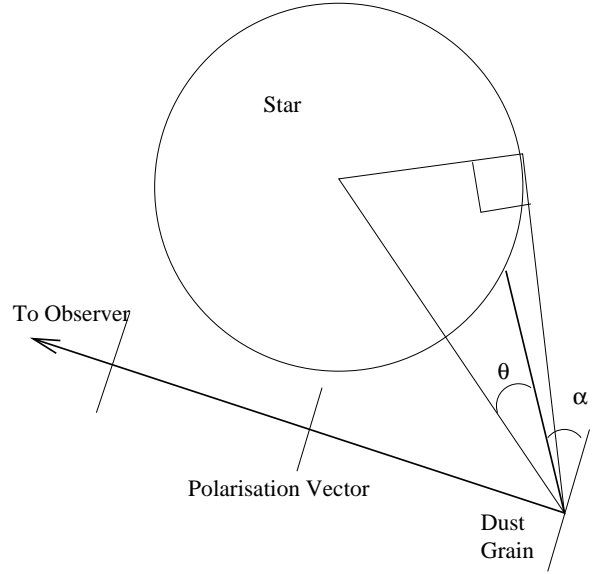}
\caption{An example of the geometry for which the polarised intensity
  was calculated, using the approximation of Rayleigh scattering from
  an optically thin shell (see text). }
\label{figGeomDia}
\end{figure}

For Rayleigh scattering, if the angle between the polarisation vector of 
the radiation seen by the observer and the ray incident on the dust grain 
is $\alpha$, then the intensity of the scattered radiation is proportional 
to $\sin^2\alpha$.
Let $\theta$ be the angle between the ray incident on the
dust grain and the direction from the dust grain to the star's centre, 
let $\phi$ be an azimuthal coordinate describing a ring on the stellar 
surface and let $\beta$ be the angle between the vector from the
star's centre to the
dust grain and the vector from the dust grain to the observer. An
example of this geometry when all vectors are co-planar is shown in
Figure~\ref{figGeomDia}.
Then, using spherical trigonometry, $\alpha$ can be expressed in
terms of $\theta$, $\phi$, and $\beta$. 
The integral that gives the scattered intensity is:

\begin{equation}
 I = \int_0^{2\pi}d\phi\int_0^{\arcsin\frac{r_s}{r_d}}
     \sin{\theta}\sin^2{\alpha}d\theta
\end{equation}

We consider the polarised
intensities of scattered light from the dust grain decomposed into
those with electric fields in radial ($I_r$) and azimuthal ($I_a$)
directions with respect to the centre of the underlying
disk. Evaluating the integral for $I_r$ and $I_a$ gives:

\begin{eqnarray}
 I_r     &=& \frac{2-u-u^2}{3} + \frac{u+u^2}{2}\cos^2\beta \\
 I_a     &=& \frac{4+u+u^2}{6}\\
{\rm where~} u &=& \sqrt{1-\frac{r_s^2}{r_d^2}}
\label{eqnScatteredI}
\end{eqnarray}

These integrals are normalised so that forward scattering from a point
source gives intensities of 1. Note that for dust distant from the
star, $u=1$ and this formula reduces to the familiar formula for
Rayleigh scattering. These intensities were numerically
integrated for the two distributions of dust discussed above to give a
two-dimensional image, and a two-dimensional Fourier transform applied
to this image to give the visibility functions arising from the dust
scattering.  Radiation back-scattered onto the surface of the star
was assumed to be absorbed. After normalising these dust
visibility functions to 1 at zero baseline, the overall visibility
functions become:

\begin{eqnarray}
 V_{\|}&=& \frac{e^{-\tau_d}V_s + (1-x_b)(1-e^{-\tau_d})V_{d\|}}
                {1 - x_b(1-e^{-\tau_d})}\\
 V_{\perp} &=& \frac{e^{-\tau_d}V_s+(1-x_b)(1-e^{-\tau_d})V_{d\perp}}
                {1 - x_b(1-e^{-\tau_d})}
\label{eqnVisFunction}
\end{eqnarray}

Here $\tau_d$ is the optical-depth of the dust, $V_{d\perp}$,
$V_{d\|}$ and $V_s$ are the visibility functions for the dust in both
polarisations and the star respectively, and $x_b$ is the fraction of
the scattered light backscattered to re-intercept the surface of the
star. The derived fit parameters for both stars are given in
Table~\ref{tblFitParams}, with the fits for both stars
shown graphically in Figure~\ref{figRCarFit}. 
In addition to quantities derived from fitting to the high-resolution
data, for comparison in Table~\ref{tblFitParams} we also present expected
stellar diameters based on the observed luminosities.
These are given in the Model~$D_s$ column, and were derived using
near-maximum dust-free M series atmospheric models from
Ireland et al. (2004a) (interpolating between the M08, M09n and M10 models), 
with the distance of the model star set to match the K band magnitudes measured 
by \citet{Whitelock00} at the same phases. Note that the J-K
colours of these stars as measured by \citet{Whitelock00} are consistent
with this model series within the observational cycle-to-cycle scatter.

\begin{table*}
 \caption{Fit parameters for both epochs of R~Car and the single epoch
 of RR~Sco, giving the apparent stellar diameter $D_s$, a model value for
 $D_s$ (see text), the diameter of the dust condensation shell $D_c$ the optical depth
 at 900\,nm $\tau_{900}$ and the reduced $\chi^2$ for the fit. Note
 that there is only one degree of freedom for the Thin Shell and
 Outflow models. Details of models given in the text.}
 \begin{tabular}{@{}llllllll@{}}
 \hline
  Star & Phase & Fit Type & $D_s$ & Model $D_s$ &$D_c$ & $\tau_{900}$ & $\chi^2$\\
       &       &          & (mas) & (mas)       &(mas)  &              &         \\
 \hline 
  R~Car & 0.08 & Thin Shell & $15.8\pm0.6$& 14.9 &$32.3\pm1.9$ & $0.19\pm0.03$ & 1.2 \\
  R~Car & 0.08 & Outflow    & $15.6\pm0.6$& 14.9 &$17.3\pm3.1$ & $0.38\pm0.07$  & 4.5 \\
  R~Car & 0.08 & UD Only    & $18.7\pm0.3$& 14.9 & - & - & 15.4\\
  R~Car & 0.15 & Thin Shell & $16.6\pm0.6$& 16.0 &$31.3\pm3.6$ & $0.14\pm0.04$ & 0.1 \\
  R~Car & 0.15 & Outflow    & $16.2\pm0.6$& 16.0 &$18.3\pm3.8$ & $0.26\pm0.06$  & 0.9 \\
  R~Car & 0.15 & UD Only    & $18.4\pm0.3$& 16.0 & - & - & 6.3\\
  RR~Sco & 0.95 & Thin Shell &$12.3\pm0.4$& 8.3 &$24.0\pm2.0$&$0.14\pm0.03$&1.1\\
  RR~Sco & 0.95 & Outflow    &$12.3\pm0.4$& 8.3 &$13.5\pm2.2$&$0.22\pm0.04$&3.7\\
  RR~Sco & 0.95 & UD Only    &$14.5\pm0.2$& 8.3 & - & - & 16\\
 \hline
 \end{tabular}
\label{tblFitParams}
\end{table*}

\section{Discussion}
\label{sectDiscuss} 

The dust free (UD only) fits in Table~\ref{tblFitParams} are clearly
eliminated for both stars.
The high $\chi^2$ values have their origin in the inability of 
this scattering-free model to generate any polarisation signal.
Of the two models with dust, the thin shell model is a better fit than 
the outflow model for both stars. The best-fit condensation radii in the
outflow models are also uncomfortably close to the continuum-forming
photospheres for both stars.
Although there may be dust types
that could plausibly exist in thermal equilibrium with the radiation
field at these radii (as discussed below), the gas temperatures in the
model star of Section~1 are 2000\,K at 1.1 stellar radii, much too
high for grain nucleation (e.g. see the discussion in
\citet{Jeong03}). The only way dust could exist so close to the
continuum-forming photosphere is if it had formed at near-minimum phases
and then fallen in. However, this kind of motion is in conflict with the
general assumptions of the outflow model.

The two epochs of R~Car do not show statistically significant
differences, but the small differences at the level of about 1-$\sigma$
are in the expected direction. 
The star appears to increase in apparent size from phase 0.08 to 0.15, 
as predicted by the models, and the optical-depth in
scattering appears to decrease, consistent with the dust being part
of an infalling layer that is partially sublimating at these
near-maximum phases. For RR~Sco, the
discrepancy between the luminosity/model-predicted
diameter and the best fit diameter is almost certainly due to
contamination from the TiO molecule. A moderate strength TiO band at
850\,nm is included in the wing of the SUSI filter profile which  
could account for the enlargement observed, although
predicting the effects of such molecular contamination is made
difficult by its strongly cycle-dependent nature (Ireland et al. 2004a).

From the discussion of dust formation radii in Section~1, it is clear 
that ``dirty'' silicates cannot form a major fraction of
the optical depth in scattering due to dust around R~Car. 
For survival within the hostile environment close to the 
photosphere, dust must have very low absorption between 1 and 
4\,$\mu$m where most of the radiation from the central star is emitted, 
and higher absorption coefficients at longer wavelengths promoting 
efficient radiative cooling. 
Corundum is a clear
possibility here, but another obvious dust type
that satisfies these criteria is forsterite, the Mg-rich olivine that
is predicted to be the first significant silicate to form in
chemical equilibrium calculations \citep{Gail99}. The optical
constants for this dust species as given in \citet{Jager03} result in
absorption coefficients 300 times higher at
10\,$\mu$m than at 2\,$\mu$m. This dust is stable at a radius of 1.2
stellar radii from the 3000\,K model star of Section~1. 
As a plausible and abundant dust candidate, we will assume that 
the dust around these stars is made of forsterite for the following
analysis of the implications of the dust model parameters.

Low absorption at
wavelengths where the bulk of the stellar radiation is emitted also
results in lower radiation pressure on the dust. Assuming full
dynamical coupling between the gas and dust and full Mg
condensation, we can calculate the radiative acceleration as a
  fraction of the gravitational acceleration for an optically-thin
  model dust shell. For a shell made of forsterite 
around the 3000\,K model star of Section~1 using Mie theory to
calculate absorption and scattering coefficients, this fraction is
41\% for spherical grains of radius 0.1\,$\mu$m and 
1.5\% for grains of radius 0.02\,$\mu$m. We therefore propose
that this dust is formed in material primarily elevated by shocks, 
which in turn means that significant temporal variation in the dust
shell optical depth and/or radius is expected.

\begin{figure}
 \includegraphics{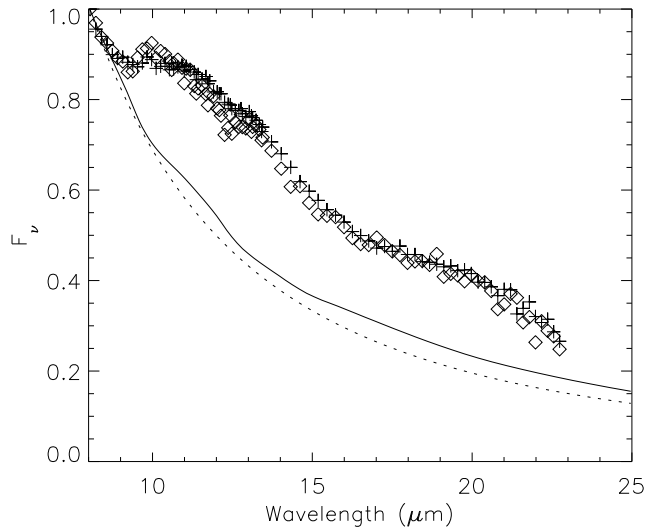}
 \caption{An example mid-IR spectrum (solid line) for a 1100\,K thin-shell model
 with a 10\,$\mu$m optical-depth of 1 at 2 stellar radii, illuminated
 by a star with a 3000\,K black-body spectrum. Flux is normalised to
 the 8\,$\mu$m flux, and the 3000\,K black-body spectrum is overlaid
 for reference (dotted line). Normalised {\em IRAS} Low Resolution
 Spectrometer (LRS) spectra are overplotted for R~Car (crosses) and RR~Sco
 (diamonds). The emission features in these spectra almost certainly
 come from a region geometrically more distant than that probed here 
 (see text).}
 \label{figMidIRSpect}
\end{figure}

Where the Rayleigh limit of small particles with respect to the 
wavelength of scattered light applies, it is sufficient to approximate
dust with a distribution of sizes as a homogeneous population with 
a single effective radius $a_e$.
For dust composed of forsterite, with a scattering optical depth
of $0.15$ at 900\,nm, typical of the observations presented here, the optical
depth in absorption at 10\,$\mu$m is 1 for $a_e$ = 55\,nm. Results
from a radiative transfer model of a geometrically thin shell of 
forsterite with a fixed temperature of 1100\,K and an
optical-depth of 1 at two stellar radii from the
centre of a spherical black-body 3000\,K star are displayed in
Figure~\ref{figMidIRSpect}. 
It can be seen that there is no strong mid-infrared feature evident
in either emission or absorption due to the dust shell.
It would seem that the dust here represents an intermediate case between
the compact geometries typically yielding absorption features, and spectra
exhibiting emission which is usually associated with extended geometries. 
This result was found to be robust, and insensitive to the optical-depth 
of the shell. Note, however, that a shell of forsterite with a 10
micron optical-depth much greater than 1 would heat up beyond
its sublimation temperature at the inner edge of the thin dust
shell. In order to meet this criteria while maintaining the observed
900\,nm optical depth, the effective particle size $a_e$ is
constrained to be larger than about 50\,nm.
From the point of view of energetics, the failure of the dust to have
a profound impact on the spectrum is simply a consequence of its optical
properties: its inability to absorb in the near-infrared means that it
cannot redistribute the bulk of the stellar flux. 
The dust scattering measured by the technique in this paper is thus 
largely independent of the form of the infrared spectrum, and could 
therefore have remained hidden from investigations based on SED
fitting. 
By this argument, the mid-infrared excess in the spectrum must come
from a material that absorbs more strongly than forsterite in the
near-infrared. Dust that has evolved and been enriched with iron
(that condenses at lower temperatures) in an outflow consisting of
shells partially ejected in previous pulsation cycles could satisfy this
property.
More detailed mid-infrared spectral modeling is beyond the scope of this
  paper, but would need to include the emission from the extended
  molecular atmosphere as predicted by self-consistent modeling by
  \citet{Jacob02} and semi-empirical modeling of \citet{Weiner04} and
  \citet{Ohnaka04a}.

The dust uncovered here is not inconsistent with the mid-infrared
  measurements of \citet{Danchi94}. These authors fit an
  outflow model to R~Leo (which has slightly more pronounced mid-infrared
  emission than either R~Car or RR~Sco), with an optical depth at
  11$\mu$m of 0.1. For forsterite dust with $a_e$ = 55\,nm, this
  corresponds to a 900\,nm optical depth of 0.025. If a similar
  outflow were to exist around RR~Sco or R~Car in addition to the
  shell model, then its effect on the fit parameters in
  Table~\ref{tblFitParams} would be roughly within the quoted
  errors. In order to measure scattering from such an outflow, shorter
  baselines and higher precision would be required.

Both R~Car and RR~Sco have relatively low mass-loss rates as constrained
by CO line observations. \citet{Groenewegen99} modeled the mass loss
rate for RR~Sco to be $1.1 \times 10^{-8}$ M$_\odot$ per year and that of
R~Car to be less than $1.6 \times 10^{-9}$ M$_\odot$ per year. If we
assume full Mg condensation at solar metallicity, then 
our best fit model with a shell at 2 stellar radii and an
optical-depth 0.15 in scattering at 900\,nm has
a total mass of $2.8 \times 10^{-6} M_\odot$ if we continue our
assumptions of dust composed of forsterite with $a_e = 0.55$\,nm
and full condensation. Less than full
condensation, or dust composed of the lower-abundance corundum would 
give an even higher mass shell. This means that the
observed mass-loss rates are only consistent with the thin-shell model
as long as the shell is considered nearly static, taking hundreds or
thousands of pulsational cycles for material in the shell to be fully
ejected. 
This is in turn consistent with the low radiative acceleration expected 
from dust that can exist so close to the stars. 

Outflow models, on the other hand, were not so easy to fit within a
self-consistent picture of stellar mass loss. Using the measured
CO outflow velocity from \citet{Young95} of 3\,km/s for RR~Sco we can
calculate the mass-loss rate from a typical outflow model in
Table~\ref{tblFitParams}. Again assuming full Mg condensation at solar
metallicity for forsterite dust with $a_e$ = 55\,nm, the mass loss
rate for a model with our best fit optical-depth 0.25 in scattering at 900\,nm and
condensation radius of 1.1 stellar radii is $2.7 \times 10^{-7}$
M$_\odot$ per year, much higher than the observed mass-loss
rates. This demonstrates once again that the outflow model can be
ruled out.

\section{Conclusion}

Using optical interferometric polarimetry, we have spatially separated the
component of flux at 900\,nm scattered by dust around the Mira
variables R~Car and RR~Sco from their photospheric emission. We found
that the inner radius of dust formation around these stars to be
  less than three stellar radii, 
consistent with dust that is relatively transparent between 1 and
4\,$\mu$m, such as iron-poor silicates or corundum. This dust exists in
a shell-like structure around these stars that may have little influence on
their mid-infrared spectra and is not part of an outflow. This
result demonstrates the complexity of the circumstellar environment
of Mira variables where mass loss and dust formation are only made
possible by pulsation. Simple outflow models are grossly inadequate,
and it is necessary to consider the changing optical properties of
dust as it evolves from the condensation radius outwards.

\section*{Acknowledgments}

Visual data for estimating the variability phases of targets at the times
of the SUSI observations were obtained from the AAVSO website. 
We thank M.~Scholz for many valuable discussions, and all the SUSI group 
members for their help in supporting the instrument. This research was 
supported by the Australian Research Council and
the Deutsche Forschungsgemeinschaft within the linkage project ``Red Giants.''

\label{lastpage}

\end{document}